\documentstyle[aps,multicol,epsfig]{revtex}

\begin{document}

\newcommand \qea {\mbox{$q_{\scriptscriptstyle {\rm EA}}$}}

\title{The Dynamics of the Frustrated Ising Lattice Gas}

\author{Jeferson J. Arenzon$^1$, F. Ricci-Tersenghi$^2$ and Daniel
        A. Stariolo$^1$}

\address{$^1$Instituto de F{\'\i}sica, Universidade Federal do Rio
         Grande do Sul\\ CP 15051, 91501-970 Porto Alegre RS, Brazil}

\address{$^2$Abdus Salam International Centre for Theoretical Physics,
         Condensed Matter Group, \\ Strada Costiera 11, P.O.Box 586,
         34100, Trieste, Italy \\ E-mails: {\tt arenzon@if.ufrgs.br,
         riccife@ictp.trieste.it, stariolo@if.ufrgs.br}}

\date{\today}
\maketitle

\begin{abstract}
The dynamical properties of a three dimensional model glass, the
Frustrated Ising Lattice Gas (FILG) are studied by Monte Carlo
simulations.  We present results of compression experiments, where the
chemical potential is either slowly or abruptly changed, as well as
simulations at constant density. One time quantities like density and
two times ones as correlations, responses and mean square
displacements are measured, and the departure from equilibrium clearly
characterized.  The aging scenario, particularly in the case of the
density autocorrelations, is reminiscent of spin glass phenomenology
with violations of the fluctuation-dissipation theorem, typical of
systems with one replica symmetry breaking. The FILG, as a valid
on-lattice model of structural glasses, can be described with tools
developed in spin glass theory and, being a finite dimensional model,
can open the way for a systematic study of activated processes in
glasses.

\end{abstract}

\begin{multicols}{2}
\narrowtext

\section{Introduction}

Upon cooling below the melting point, liquids may either crystallize
or enter a super cooled regime. In the latter case, as the glass
transition temperature $T_g$ is approached, molecular motion gets
slower and slower and the viscosity increases enormously. The
relaxation time increases by several orders of magnitude and for all
practical purposes the system remains out of equilibrium. Although
mechanically responding as a solid, structural relaxation is still
present, slowing down as the system ages. The response to a
perturbation applied at a particular time $t_w$ will persist for very
long times (long term memory), preventing the system from reaching
equilibrium.  While the system ages, one time quantities
asymptotically tend to their equilibrium values while two times
quantities depend explicitly both on the observation time and on the
time when the perturbation was applied: time translation invariance
(TTI) is broken, which is a manifestation of history dependence. Upon
cooling the system gets trapped on long lived meta stable states which
depend on the cooling rate, eventually escaping as a result of
activated processes.  This complex glassy dynamics is reflected in
several characteristic features, as non-exponential relaxation,
breakdown of fluctuation-dissipation relations, etc, and there is a
widespread believe that the underlying rugged energy landscape plays a
fundamental role on the dynamics\cite{landscape}.  Many aspects
related to the structural glass transition and the nature of the
glassy phase are still poorly understood~\cite{science,angell}, 
as a consequence of the inherent complexity of the underlying physics
and the corresponding lack of a simple but non-trivial model.  Only
recently some analytical progress has been made with classic models
based on Lennard-Jones or soft spheres potentials~\cite{barbara}
which, on the other hand, have been extensively studied via molecular
dynamics simulations~\cite{kob}.  Meanwhile, in the field of spin
glasses, disordered magnetic systems which share many physical
properties with structural glasses, a reasonable theoretical
understanding of the basic physics has been achieved at least at the
mean field level~\cite{fischer,young}.  Recently it has been found
that the equations describing dynamical correlation functions of a
kind of mean field spin glasses simplify, above the transition, to the
single equation of the Mode Coupling Theory of super cooled
liquids~\cite{gotze}. In particular, several results point out that
these models (e.g.\ the $p$-spin model) with one step of replica
symmetry breaking and structural glasses are in the same universality
class~\cite{kirkpatrick,cuku}, indicating a deeper analogy between the
physics of spin glasses and structural glasses than previously
thought. Nevertheless, it is still matter of debate to what extent
this analogy can be pushed forward. Some important differences between
these two kinds of systems are evident: while a defining feature of
spin glasses is the quenched disorder and frustration, no obvious
quench disorder is present in structural glasses. Several models with
frustration but no disorder have been studied recently~\cite{nodis}.
Another important difference is the fact that the dynamical variables
in a spin glass are localized in space and do not diffuse as the
molecules in a super cooled liquid. We would like to have at hand a
simple, yet non trivial, microscopic model of super cooled liquids in
which the fruitful techniques developed in spin glass theory could be
applied and possibly be extended to finite dimension. Moreover, Mode
Coupling Theory and mean field models in general do not describe
activated processes, which are responsible for some important
characteristics of the glass phase, as e.g.\ cooling rate dependence
of the asymptotic state.  There are some other very simple models with
kinetic constraints which reproduce quite well the glassy
phenomenology~\cite{jackle,ka}, but being essentially dynamical they
are hard to be studied with analytic techniques.

In this paper we describe salient dynamical features of a Hamiltonian
lattice model of structural glasses, the Frustrated Ising Lattice Gas
(FILG). The interesting properties revealed by the model rest on an
interplay between two different kinds of degrees of freedom: diffusive
particles or translational degrees of freedom and internal or
orientational degrees of freedom. We present results of Monte Carlo
simulations of one and two times quantities, like density relaxation,
density-density correlations and responses, correlations and responses
of internal degrees of freedom, mean square displacements of the
particles and compression experiments. This three dimensional model
may serve as a numerical laboratory to perform a systematic study of
the role played by activated processes on the long time dynamics of
glass formers, a feature that is missing from mean field
approximations.

The paper is organized as follows. In section~\ref{section.filg} we
review the essential results both from equilibrium and out of
equilibrium dynamics for the FILG. In section~\ref{section.cycle}
results for slowly varying chemical potentials are presented and
compared with the case where a sudden quench is performed, while in
section~\ref{section.fdt} the breakdown of the fluctuation-dissipation
theorem in this model is analyzed. Finally, in
section~\ref{section.conclusions}, we present some conclusions and
discuss some open questions.

\section{The Frustrated Ising Lattice Gas}
\label{section.filg}

The FILG~\cite{coniglio94} is defined by the Hamiltonian:
\begin{equation}
{\cal H} = -J \sum_{<ij>} (\varepsilon_{ij} S_i S_j - 1)n_i n_j - \mu
\sum_i n_i .
\label{H}
\end{equation}
The dynamical variables $n_i=0,1$ ($i=1\ldots N$) are local densities
or site occupations. The $S_i$ represent internal degrees of freedom,
e.g.\ rotational ones. Although molecules in glass forming liquids may
assume several spatial orientations, here we take the simpler case of
only two possibilities, $S_i=\pm 1$. The usually complex spatial
structure of these molecules is in part responsible for the geometric
constraints, imposed by the neighborhood, on their translational and
rotational dynamics.  This hindrance effect is mimicked by the
quenched random variables $\varepsilon_{ij}=\pm 1$. When $J\rightarrow
\infty$, in order to minimize ${\cal H}$, either the spins should
satisfy the bond $\varepsilon_{ij}$ or at least one of the sites $i$
or $j$ must be empty.  In this limit, the (site) Frustrated
Percolation~\cite{coniglio94} model is recovered, where no frustrated
link can be fully occupied, implying that any frustrated loop in the
lattice will have a hole and then $\rho<1$, preventing the system from
reaching the close packed configuration.

The equilibrium properties of the $3D$ model have been studied
in~\cite{nicodemi} for $T=1$ and $J=10$, which corresponds in practice
to the Frustrated Percolation limit.  In the low density regime ($\mu
\leq 0.75$) the behavior is liquid like, time correlation functions
decay exponentially, equilibration is quickly achieved and the
particles mean squared displacement grows linearly with time, a simple
diffusion scenario since particles hardly feel any constraint in their
mobility.  At $\mu \approx 0.75$ there is a percolation transition,
the corresponding density being $\rho \approx 0.38$.  This first
transition manifests dynamically in the onset of two different
relaxation regimes in the correlation functions, a fast exponential
relaxation at short times and a slow, stretched exponential relaxation
at longer times. While for intermediate times the diffusion is
anomalous, for longer times it is linear with the diffusion
coefficient becoming smaller as the density grows with growing
chemical potential (or equivalently, with lowering temperature). A
second transition appears for $\mu_c \simeq 5.5$, corresponding to a
density $\rho_c \simeq 0.67$. This is a glass transition at which the
relaxation times diverge and the diffusion constant goes to zero. In
the FILG the structural manifestation of this transition is the
presence of a {\it frozen} percolating cluster. This transition
corresponds to the dynamical one in mean field $p$-spin or Potts
glasses or the ideal glass transition in the mode coupling theory.
Scarpetta {\it et al}~\cite{SFP} studied an equivalent, non local
version of the FILG (the Site Frustrated Percolation, SFP) in 2D,
finding a behavior similar to the 3D FILG, the main difference being
the (Arrhenius) dynamical singularity occurring at zero temperature.
It was also found in~\cite{nicodemi} that the spin glass
susceptibility $\chi_{\scriptscriptstyle SG}$, associated with the
internal degrees of freedom of occupied sites, diverges at the same
value of $\mu$ where a glass transition takes place. This is a
thermodynamic spin glass transition associated with the frozen-in of
the internal degrees of freedom. On the other hand, the
compressibility associated with the density variables does not present
critical behavior but only a maximum as observed in many glass
formers. In view of this behavior one could ask whether a suitably
defined nonlinear compressibility should diverge at the glass
transition.  This should indicate the thermodynamic Al character of the
glass transition in this model. A nonlinear compressibility,
$\kappa_{\scriptscriptstyle SG}=N(<q_n^2>- <q_n>^2)$, can be
introduced with $q_n=N^{-1}\sum_i n_i^1 n_i^2$ where 1 and 2 label two
real replicas of the system.  One could think that in this quantity,
associated with fluctuations of an Edwards-Anderson like order
parameter involving only density variables, a divergence analogous to
the one present in the $\chi_{\scriptscriptstyle SG}$ might be
detected.  This quantity has been measured~\cite{notpub} but no
evidence of critical behavior has been found near $\mu_c$:
$\kappa_{\scriptscriptstyle SG}$ attains a plateau that seems to be
roughly sample and size independent. The fact that the spin variables
present a thermodynamic transition while the particles do not points
once more to the purely dynamical character of the glass
transition. Of course, a thermodynamic transition may be present at a
lower temperature (or a higher chemical potential) corresponding to
the Kauzmann $T_K$ and associated with a configurational entropy
collapse, but this remains an open question in this model which
deserves further study.

At a mean field level, equilibrium properties have been studied in
closely related models~\cite{jef}. For large $\mu$ all sites are
occupied and the behavior corresponds to the Sherrington-Kirkpatrick
model, with a continuous spin glass transition. Upon decreasing $\mu$
one reaches a tricritical point, below which the transition is first
order. In this region, analogously to other first order transition
models like the $p$-spin and the Potts glass~\cite{potts}, above the
critical temperature there is a dynamical transition where meta stable
states first appear\cite{jef}. We still do not know to what extension
this scenario remains valid in finite dimensions.

A characteristic feature of the out of equilibrium dynamics of glassy
systems is aging~\cite{aging}. Slow aging dynamics is present in the
FILG both in density-density two times correlations as well as in
correlations of the internal degrees of freedom~\cite{pre} after
performing sudden quenches in $\mu$ from a very small value
characteristic of the liquid phase (low density) to a high value
corresponding to the glassy phase. Performing a quench in $\mu$ is
similar to the application of a sudden compression, we raise the
density of the system keeping the ratio $\mu/J \leq 1$ in order to
prevent close packing. Connected two point correlations can be defined
as:
\begin{equation}
c(t,t_w) = \frac{1}{N} \sum_i n_i(t) n_i(t_w) - \rho(t) \rho(t_w) ,
\label{density-density}
\end{equation}
where the global density at time $t$ is given by $\rho(t)=N^{-1}\sum_i
n_i(t)$. We now define the density autocorrelations as $C_n(t,t_w) =
c(t,t_w)/c(0,t_w)$.

\begin{figure}
\centerline{\epsfig{file=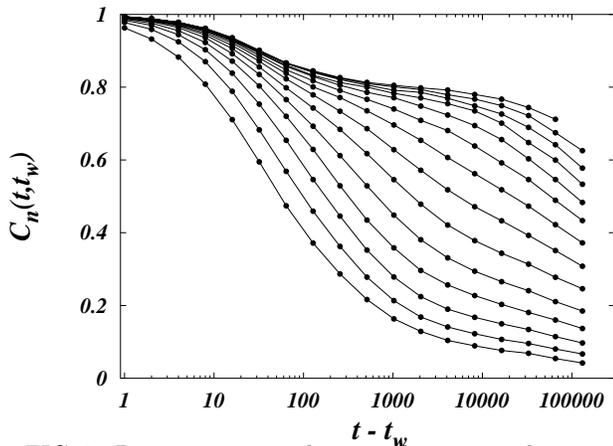,height=\columnwidth,angle=270}}
\caption{Density autocorrelations after a quench to $\mu=10$ at $T=1$
and $J=10$ for $L=20$. The waiting times range from $2^5$ (bottom) to
$2^{17}$ (top) and the averages are over 50 samples.}
\label{densitycorrelation}
\end{figure}

In Fig.~\ref{densitycorrelation} the behavior of $C_n(t,t_w)$ is shown
as a function of $t$ in a semi log plot after a quench in chemical
potential to a value $\mu=10$, for waiting times between $2^5$ and
$2^{17}$. A typical aging scenario is present signalling the slowing
down of the dynamics as the waiting time grows. For the longest
waiting times the correlation presents a rather fast relaxation to a
plateau in which the system evolves in quasi equilibrium: the dynamics
is stationary and the fluctuation-dissipation relations hold.  The
plateau separates two time scales typical of glassy systems: a $\beta$
(fast) relaxation for short times and an $\alpha$ (slow) relaxation at
longer times, corresponding respectively to the fast movements of the
particles inside the dynamical cages and the large scale, cooperative
process that takes much more time in order to rearrange the cages.
Moreover, in this very long time regime ($t-t_w \gg t_w$), the system
falls out of equilibrium, the correlations decay to zero
asymptotically and time translational invariance (TTI) no longer holds
with the corresponding violation of the fluctuation-dissipation
theorem (FDT), as will be shown in the following sections.  A good
scaling of the correlation functions is obtained assuming a time
dependence of the form $h(t_w)/h(t+t_w)$ with $h(x)$ given by
$h(x)=\exp\left[(1-\alpha)^{-1}(x/\tau)^{1-\alpha}\right]$ where
$\alpha < 1$ and $\tau$ is a microscopic time scale~\cite{pre}. This
form is quite general as one recovers the cases of a simple $t/t_w$
dependence (full aging) when $\alpha=1$ and stationary dynamics when
$\alpha=0$~\cite{aging}. A similar aging scenario has been found
associated with the relaxation of internal degrees of
freedom~\cite{pre}.

\section{Slow and Fast Compressions}
\label{section.cycle}

\begin{figure}
\centerline{\epsfig{file=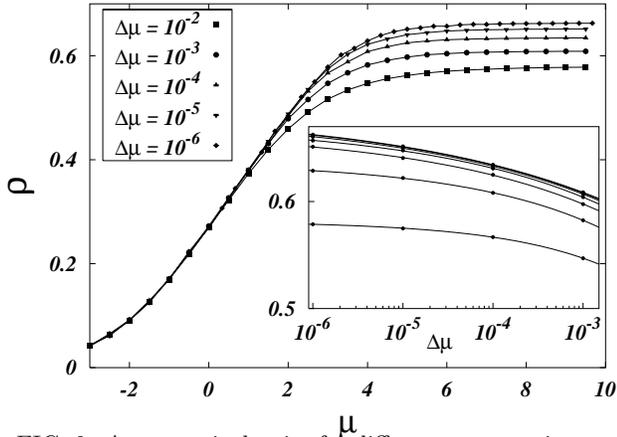,
height=\columnwidth,angle=270}}
\caption{Asymptotic density for different compression rates, ranging
from $10^{-2}$ to $10^{-6}$ (bottom to top).  Inset: power law fitting
for several values of $\mu$.}
\label{annealing}
\end{figure}

When the system is gently compressed the glass transition
depends on the rate of compression . For a fixed temperature we have done simulations of slow
compression experiments, where at each Monte Carlo step (MCS) $\mu$ is
increased by a quantity $\Delta\mu$. Fig.~\ref{annealing} shows the
evolution of the density for several compression rates, ranging from
$10^{-2}$ to $10^{-6}$.  It can be noted that for slower compressions,
the final density is higher. Up to $\mu \approx 1$ the density remains small enough and the system is able to
equilibrate sufficiently fast, regardless of $\Delta\mu$, and all curves
collapse. As the density increases, the equilibration time also
increases since the system cooperative rearrangement involves a larger
number of particles in a smaller region. In analogy with cooling
experiments, for each compression rate the system gets out of
equilibrium at a different value of $\mu$ and the asymptotic density
increases with decreasing rates. The points where the curves for each
rate depart from the master curve define the glass transition point
$\mu_g$. The dependence of $\mu_g$ on the compression rate is
characteristic of glass formers and is thought to be consequence of
activated processes: no cooling (compression) rate dependence is
observed in mean field models where the glass transition appears at a
fixed temperature (chemical potential).  For a fixed $\mu$, by
decreasing the compression rate, the density $\rho(\mu)$ tends to the
equilibrium density $\rho_{\infty}(\mu)$ as a power law
$\rho_{\infty}(\mu)-\rho(\mu) \propto \Delta\mu^{-b}$, as can be seen
in the inset of Fig.~\ref{annealing}, the parameters depending on the
value of $\mu$. The extrapolated equilibrium value is slightly
higher than the one obtained upon fast compressions, as will be
shown later.

\begin{figure}
\centerline{\epsfig{file=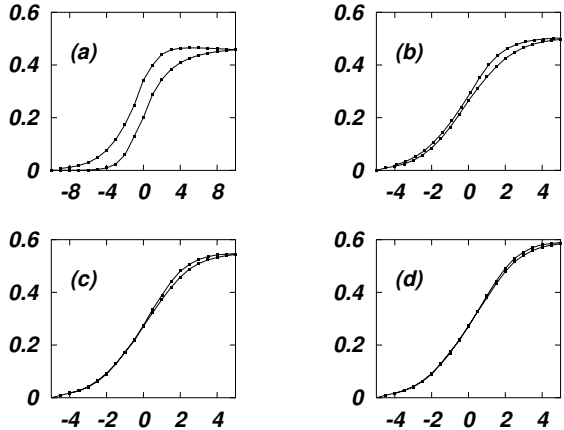,height=\columnwidth,angle=270}}
\caption{Hysteresis curves ($\rho \,vs \mu)$ from cycling experiments in a $L=6$ system
($T=1$ and $J=10$). Notice the different $x$-axis scale of
fig.$(a)$. The cycling rates are, from $(a)$ do $(d)$, 1, $10^{-1}$,
$10^{-2}$ and $10^{-3}$.}
\label{cycle}
\end{figure}

In analogy with temperature
cycling experiments in spin glasses which show the onset of irreversible
processes, here, by increasing and decreasing the chemical potential
at a given rate, hysteresis cycles can be produced whose internal area
increases with $\Delta\mu$. Moreover, the chemical potential range in
which the system behaves in an irreversible way also increases with
$\Delta\mu$. In Fig.~\ref{cycle} the hysteresis curves are presented for
several cycling rates.  Notice that the lower left part of the cycle
coincides with the equilibrium curve and for very slow rates the cycle
tends to disappear and we recover the above annealing equilibrium
curve.  As before, the maximum achieved density increases for slower
rates.

\begin{figure}
\centerline{\epsfig{file=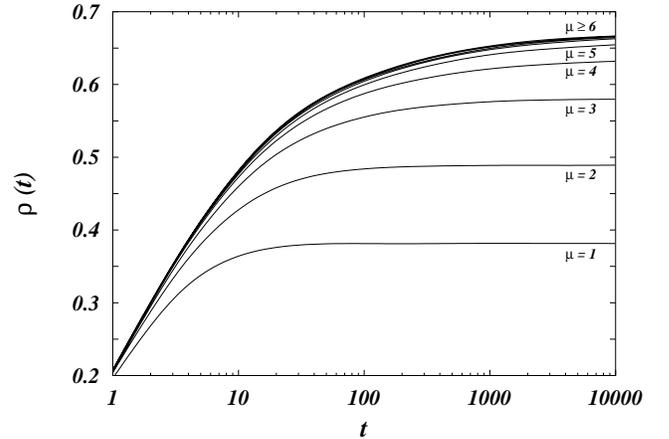,width=\columnwidth,angle=0}}
\caption{Increasing of the density in a $L=40$ system after a sudden
quench in the chemical potential ($T=1$ and $J=10$). All the data for
$\mu \ge 6$ collapse on the same curve, which can be well fitted by a
power law (see text).}
\label{dens}
\end{figure}

\begin{figure}
\centerline{\epsfig{file=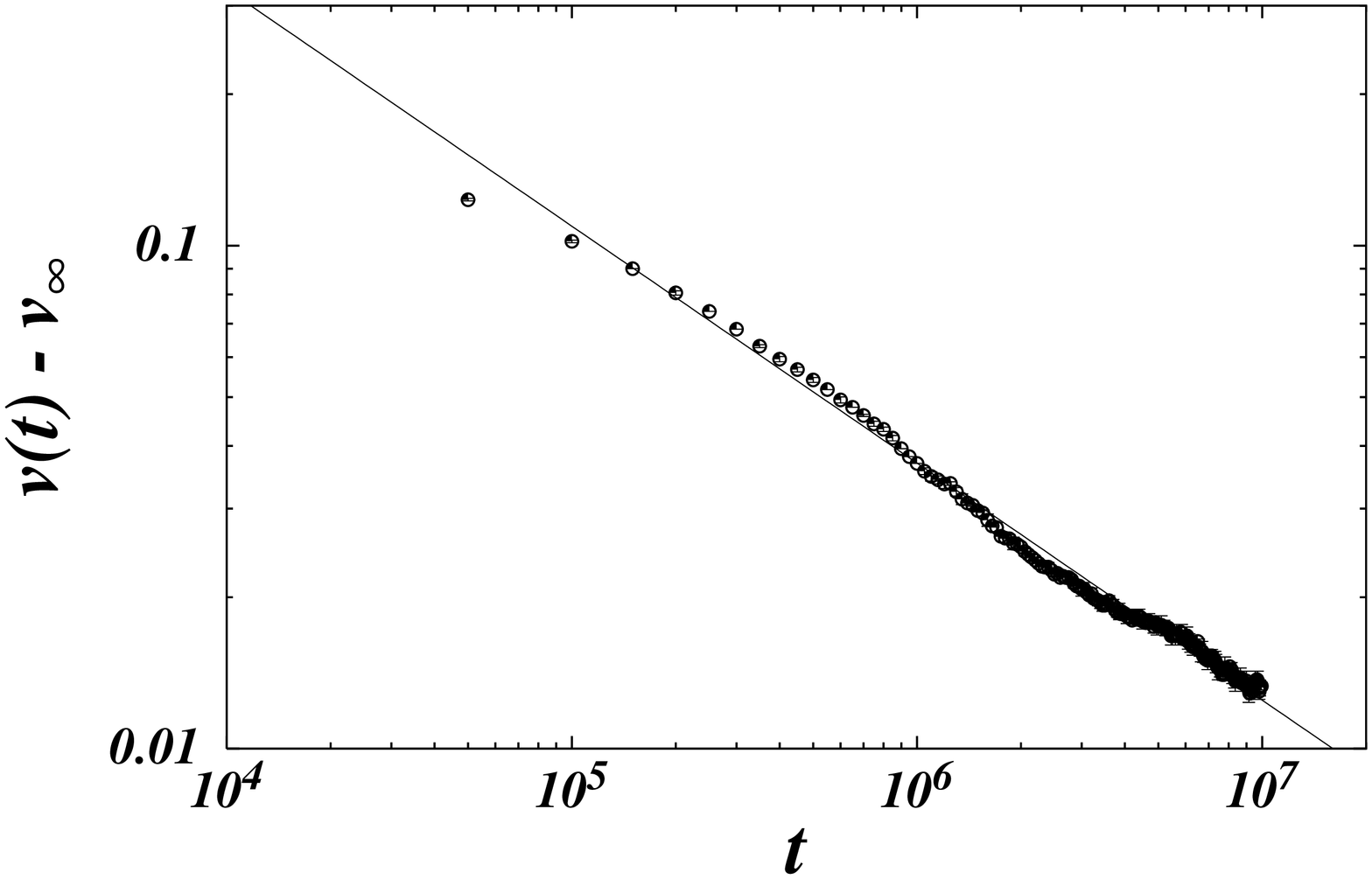,height=\columnwidth,angle=270}}
\caption{Decay of the specific volume after a quench to $\mu=10$ for a
lattice with $L=6$. The solid line is a power law fit over the last
two decades, $v(t)=v_{\infty} + a t^{-\alpha}$, with $\alpha \simeq
0.47$ and $v_{\infty} = 1/ \rho_{\infty} \simeq 1.48$ ($\rho_{\infty}
\simeq 0.673$).}
\label{density}
\end{figure}

Instead of slowly compressing the sample, high densities can also be
obtained by a fast compression analogous to temperature quenching
experiments. Here we consider an initial state with an empty lattice,
corresponding to $\mu = -\infty$.  Then the chemical potential is
suddenly increased to a value above the critical one.  In
Fig.~\ref{dens} we show how the density increases after a sudden
quench to different $\mu$ values, ranging from 1 to 10. For all $\mu <
\mu_c \simeq 5.5$ the density converges more or less rapidly to an
asymptotic value, which depends on $\mu$.  For $\mu > \mu_c$ the
density increases with a law $\rho(t)$ which seems to be independent
of the chemical potential and that converges to the asymptotic value
with a power law: $\rho_\infty-\rho(t) \propto t^{-\alpha}$. The best
value for the parameters are $\alpha \simeq 0.47$ and $\rho_\infty
\simeq 0.673$, slightly lower than the value obtained in the slow
compressions. In contrast to a sudden compression, by gently
annealing the system can approach equilibrium more efficiently.
Nevertheless one can expect that the equilibrium density will be
reached after a quench in a much longer time scale driven by activated
processes. In Fig.~\ref{density} we show the relaxation of the density
after a sudden compression to $\mu=10$, deep in the glass phase.

If we assume that the density asymptotic value is essentially
determined by the amount of frustration present in a sample, then it
is possible to calculate the range of allowed values for
$\rho_\infty$.  In a sample with no frustration obviously
$\rho_\infty^{\rm max} = 1$, while in a fully frustrated sample we
have that $\rho_\infty^{\rm min} = 5/8 = 0.625$. A fully frustrated
sample has the interactions $\varepsilon_{ij}$ such that every
plaquette is frustrated.  The value we find for the random model,
$\rho_{\infty} \simeq 0.673$, falls in this interval, as it should,
and is not very far from the one found in the fully frustrated case.
This would suggest that the FILG defined on a fully frustrated lattice
may be an interesting non random model with features similar to those
presented in this paper. In fact we believe that the glassy behavior
of the FILG is based much more on its frustration rather than on its
randomness.

As long as activated processes do not enter the scene we would expect
a behavior reminiscent of mean field models.  In the thermodynamic
limit, mean field theory for the $p$-spin spin glass predicts that the
internal energy will relax with a power law to a threshold value
$E_{th}$ greater than the equilibrium one~\cite{cuku}. The absence of
activated processes will prevent the system from escaping the
threshold states and it will be kept out of equilibrium forever. In a
true glass the relaxation of the density at sufficiently low
temperatures should behave this way. In the case of the FILG we do not
know what the equilibrium density is, so in order to test the previous
hypothesis we have measured autocorrelations at constant density.

\begin{figure}
\centerline{\epsfig{file=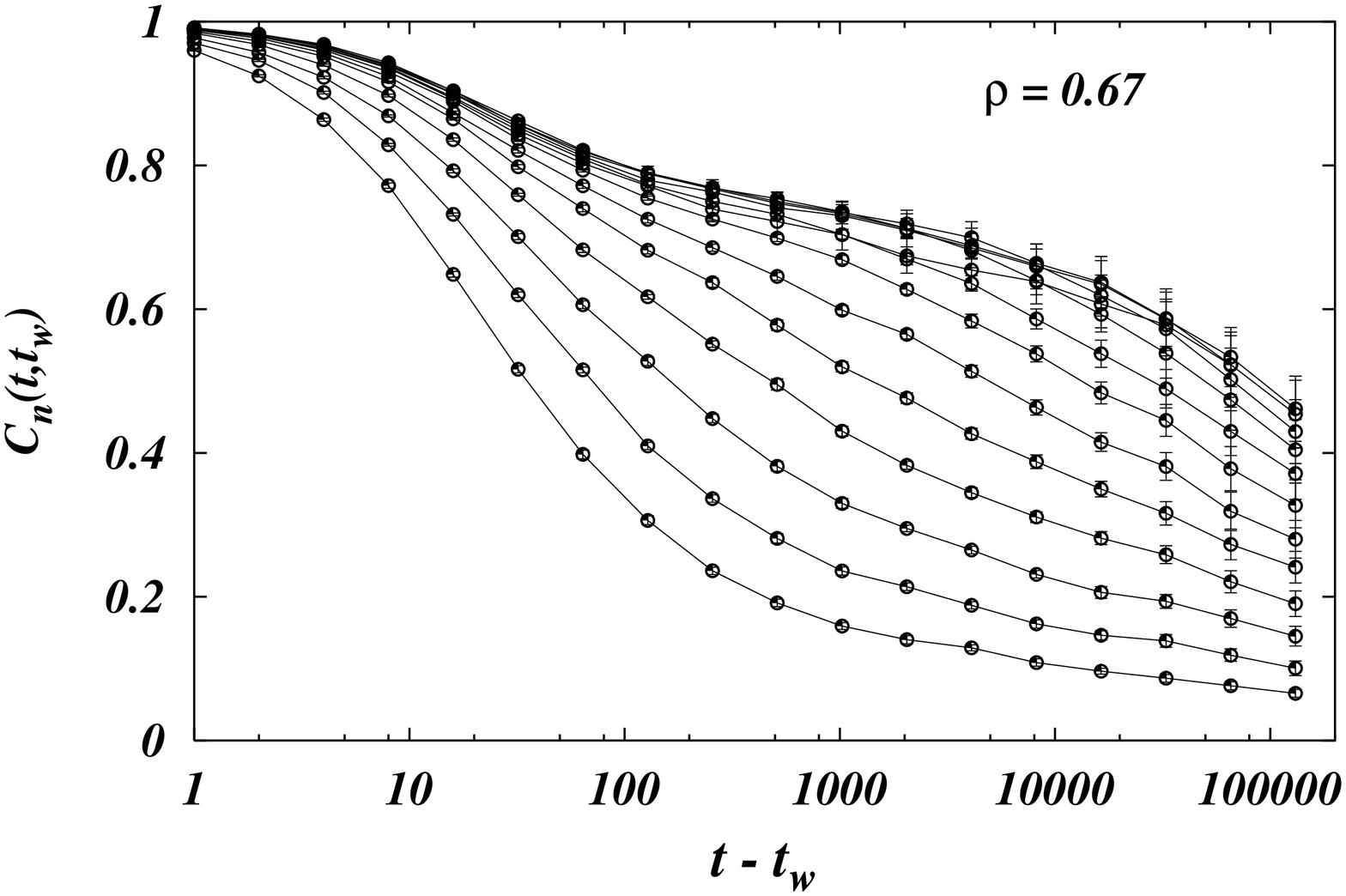,height=\columnwidth,angle=270}}
\centerline{\epsfig{file=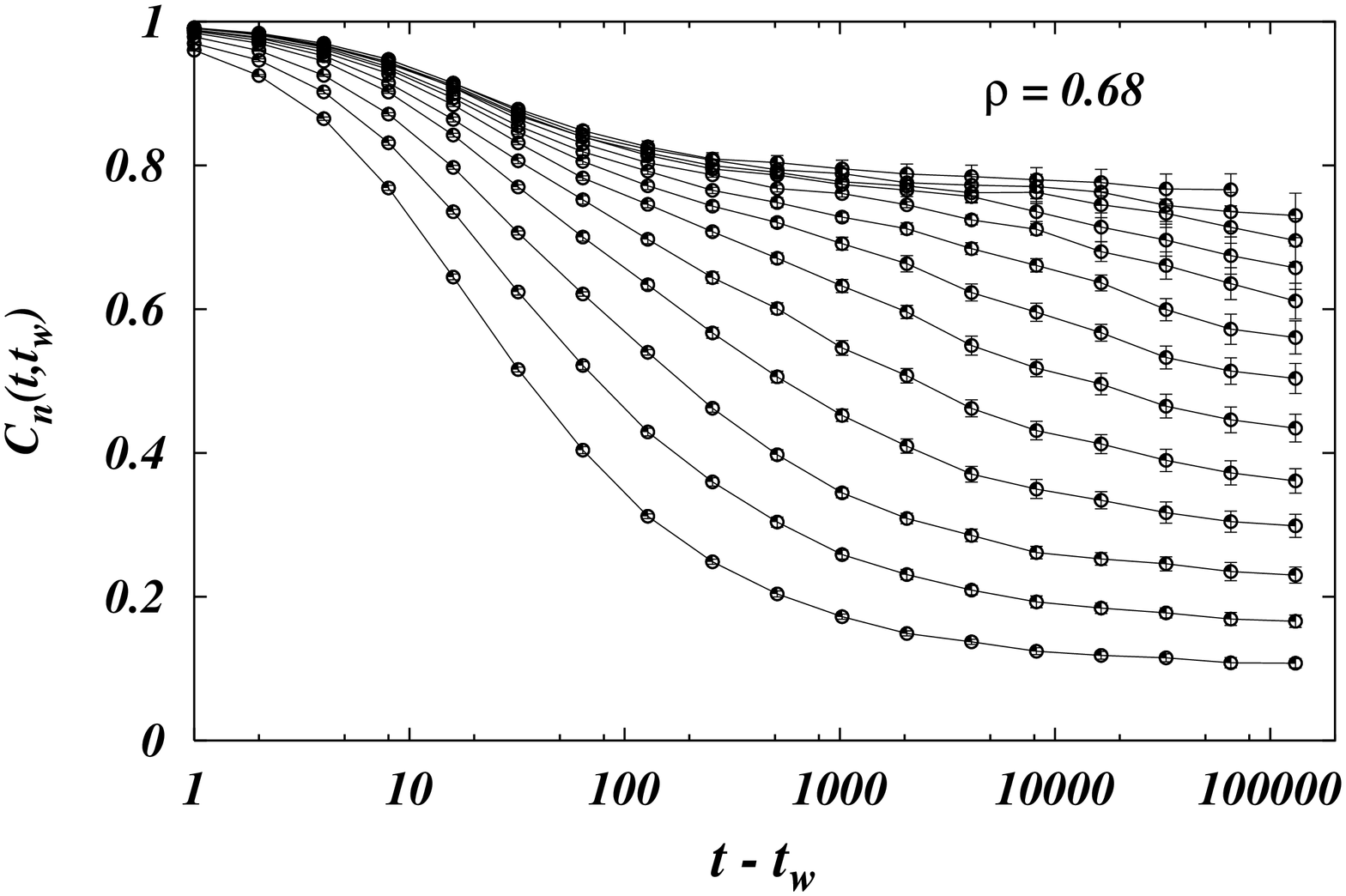,height=\columnwidth,angle=270}}
\caption{Decay of the density autocorrelations after a quench at a
constant density and $L=20$. Note that for $\rho=0.67$ (top) and
moderate waiting times the dynamics becomes stationary and aging
stops, quite different from what happens for $\rho=0.68$ (bottom).}
\label{constant} 
\end{figure}

Considering again the connected two times autocorrelations
(eq.~\ref{density-density}), we present in Fig.~\ref{constant} results
of simulations of quenches at fixed densities.  The upper figure shows
results for a density that is slightly below the critical one
($\rho=0.67$) for several waiting times $t_w$.  There exists a
characteristic waiting time after which all the curves collapse on a
master one, meaning that aging is interrupted and time translation
invariance (TTI) is recovered. The relaxation preserves the
characteristic two steps with equilibrium dynamics for large $t_w$.
On the other hand, for a density $\rho=0.68$, slightly above the
critical one, the system seems to age forever, as can be seen in the
lower figure. In this case, even for the largest waiting times, the
quasi-equilibrium regime is followed by an out of equilibrium
relaxation where TTI no longer holds and fluctuation-dissipation
relations are violated. The behavior in this two cases is
qualitatively different and is reminiscent of mean field behavior with
a threshold or critical value for the density. This will be reflected
also in the form in which the fluctuation-dissipation theorem is
violated, as will be discussed below.

\begin{figure}
\centerline{\epsfig{file=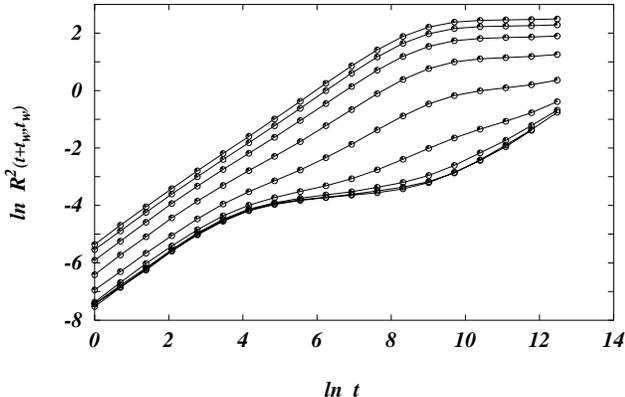,height=\columnwidth,angle=270}}
\caption{Two-times mean square displacement after a quench for $J=10$,
$T=1$ and $L=20$. From top to bottom: smaller to higher $t_w$.}
\label{diffusion}
\end{figure}

In real space, one way of seeing aging effects is looking at the
diffusion properties of the particles.  The averaged mean square
displacement (MSD) in the time interval $(t,t_w)$ is defined by
\begin{equation}
R^2(t,t_w)=\frac{1}{N}\sum_i a_i(t,t_w)
\left[ \vec{r}_i(t) -\vec{r}_i(t_w)\right]^2
\end{equation}
where the function $a_i(t,t_w)$ is 1 if the $i$-$th$ particle is
present at both times $t$ and $t_w$ and is 0 otherwise. In this way
only particles that do not leave the system enter the
measures. Obviously, if the density is kept constant, $a_i=1 \;
\forall i$. For the smallest waiting times, the density is low and
diffusion is like in a simple fluid, $R^2$ being linear in $t$ (see
Fig.~\ref{diffusion}). Only
for large times the density attains a large value, cages are formed
trapping the particles, and the diffusion is arrested, as signaled by
the appearance of a plateau. For larger $t_w$ and $t \ll t_w$, the
particles are rattling inside the cages formed by their neighbors and
the diffusion is still normal, $R^2$ being again linear in $t$. The
fact that the first linear regime is longer the smaller is $t_w$ is
simply because for larger $t_w$ the density at $t=0$ is larger. As
soon as $R^2$ is comparable to the cage linear size, the diffusion
halts and a plateau is attained. To escape from the cage (hence from
the plateau), a large, cooperative rearrangement must occur among
neighbor cages.  This slow phenomenon is reflected in the extension of
the plateau. Moreover, for larger times we are essentially measuring
the diffusion of the cages, what can be seen as the second linear
regime present in Fig.~\ref{diffusion}. From this figure it is clear
that aging effects with characteristic dynamical regimes are also
present in real space properties of the system.

\section{FDT}
\label{section.fdt}

When glasses evolve slowly out of equilibrium fluctuation-dissipation
relations are no longer valid. Instead, in computer simulations and
mean field approximations a generalized relation is
obtained~\cite{cuku,aging,fdt}:
\begin{equation}
R(t,t_w) = \frac{X[C(t,t_w)]}{T}\frac{\partial C(t,t_w)}{\partial t_w}\ ,
\label{gfdt}
\end{equation}
where $C(t,t_w)$ is a two times correlation function and $R(t,t_w)$
with $t\!>\!t_w$ is the associated response. $T$ is the heat bath
temperature and $X(C)$, the ``fluctuation-dissipation ratio''(FDR), is
a function that measures the departure from FDT: at equilibrium $X=1$
and the usual FDT is recovered, while in the out of equilibrium
regime, $X<1$. The function $X$ turns out to be a constant in mean
field models of glasses, signalling that the asymptotic dynamics of
the system is dominated by only one time scale (apart from the FDT
scale).  In this time scale, corresponding to the aging regime, the
slow degrees of freedom are at an effective temperature different from
the bath one and that depends on the FDR as
$T_{eff}=T/X$~\cite{cukupe}. Recently the first attempt at an
experimental measurement of this effective temperature has been done
in glycerol glass~\cite{grigera}.

In a finite system aging will eventually be interrupted due to
activated processes leading the system to equilibration at the
temperature of the bath.  Nevertheless the true role of activated
processes in the long time dynamics of glassy systems still remains to
be understood. While an analytic approach to the problem is very
difficult one may gain insight from simulations on finite systems.

\begin{figure}
\centerline{\epsfig{file=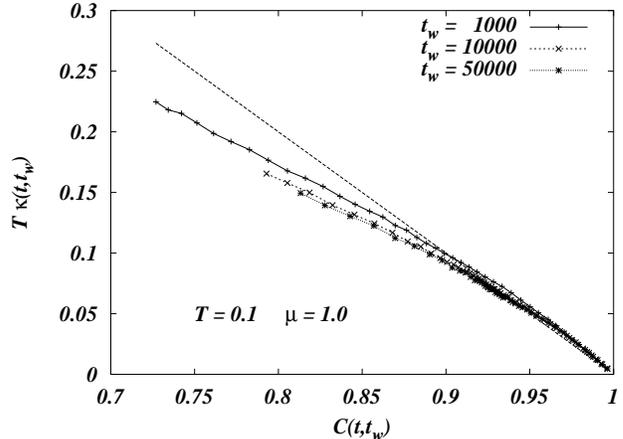,width=\columnwidth,angle=0}}
\caption{Integrated response versus density correlations for three
waiting times (see text).}
\label{fdt_1}
\end{figure}

We have measured the FDR on the FILG~\cite{letter}. The main result is
shown in Fig.~\ref{fdt_1}, where $X$ is given by minus the derivative
of the data curves. In this case we have measured density correlations
and associated compressibilities (for details see~\cite{letter}).
This result was obtained in a lattice of linear size $L=30$.
The result is the one expected for a model having only two
dominating time scales: the FDT one, when $C > \qea$ and another one,
when the system is relaxing very slowly out of equilibrium and $C <
\qea$, \qea being the Edwards-Anderson (EA) order
parameter~\cite{fischer}. The constancy of $X$ in the aging regime is
exactly what is predicted by mean field theory.

What is very remarkable is that this constancy seems to be valid not
only for the asymptotic curve ($t_w \to \infty$), but also for
finite (although large) $t_w$ curves.  Then we can express the FDR as 
function only of the smaller time, $t_w$, and we observe that it changes very
slowly with $t_w$.

We also note that the fact that $X$ seems to depend only on the
smaller time is in agreement with some recent analytical results.
In ~\cite{fravi} it has been shown that slow glassy dynamics can be
viewed as an evolution between ``quasi states'', where the system
(almost) equilibrates before relaxing further.  When the field is
switched on at time $t_w$ it will select a particular set of
quasi-states, those with the ``right'' magnetization. If this is the
case then the externally induced small perturbation~\cite{note1} will
regress in the same way a spontaneous fluctuation does and one can
show~\cite{silvio} that the FDR may depend on the smaller time, but
not on the larger one. And this is exactly what we have observed
numerically in the FILG model.

Nevertheless, in contrast with mean field predictions, the curves for
different temperatures and chemical potentials in the glass phase (for
a fixed large $t_w$) remain parallel, simply shifting upwards with a
corresponding decrease in the value of the EA parameter~\cite{letter}.
This is reasonable as one expects that the FDT regime will be the only
one present as $T \rightarrow T_g$.  This behavior is not observed in
mean field approximation, where the EA parameter has a discontinuous
jump at $T_g$ and $X$ changes continuously for $T < T_g$. Other
studies in long range models seem to point out a continuous approach
of $T_{eff}$ to the bath one~\cite{felix}. A systematic study of these
effects is necessary in order to understand the underlying mechanisms
responsible for them.

It is not clear whether the system will undergo also a static transition
with one step of replica symmetry breaking, as the off-equilibrium
dynamics would suggest.  Thermalization at $T\leq T_g$ is really hard
to attain and so we did not succeed in answering this question.
However, even in the case that all the observed dynamical features
would not correspond to an underlying static transition and they would
eventually disappear as the system equilibrates, the results are
interesting for at least two reasons.  First, real glassy systems can
be in the off-equilibrium regime all along their life and so it is
more relevant the study of their out of equilibrium behavior than the
static one.  Second, even if what we are observing is due to finite
time effects (although very large), it is very remarkable that it is
well described by mean field approximations.  Maybe is the case that
on time scales where activated processes are very rare mean field
theories (that completely neglect activated processes) are really a
good approximation.

\begin{figure}
\centerline{\epsfig{file=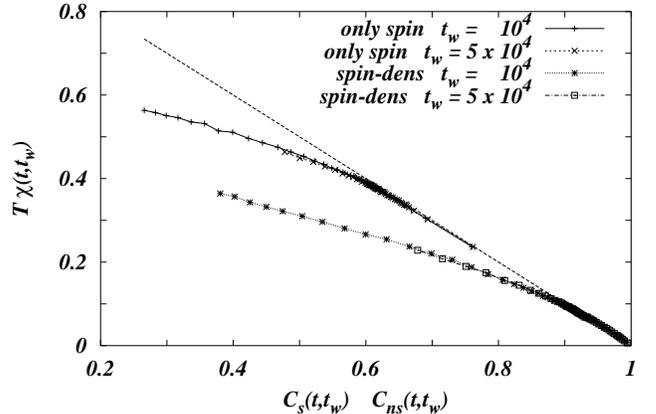,width=\columnwidth,angle=0}}
\caption{Integrated response versus spin correlations (see text).}
\label{fdt_2}
\end{figure}

We have also measured the FDR associated with the relaxation of the
spin or internal degrees of freedom. One can consider the spin-spin
autocorrelation
\begin{equation}
C_s(t,t_w) = \frac{1}{N} \sum_i S_i(t)S_i(t_w)
\end{equation}
or, only taking spins from occupied sites, the
spin-density-spin-density one
\begin{equation}
C_{ns}(t,t_w) = \frac{1}{N} \sum_i S_i(t)n_i(t)S_i(t_w)n_i(t_w)
\end{equation}
By applying suitable perturbations, like small random magnetic fields,
one can define in the standard way the integrated responses
corresponding to the correlations defined above. The parametric plot
of the integrated responses versus correlations is shown in
Fig.~\ref{fdt_2}.

The interesting result which shows up here is that, qualitatively, the
FDR is the same in the three cases. Furthermore, fits to the data show
approximately the same asymptotic value of $X$. While the data for
spin-density-spin-density and density-density are exactly on the same
straight curve the data for spin-spin variables depart from the FDT
line at a smaller value of the Edwards-Anderson parameter. This is due
to the fraction of spins placed on empty sites.

The fact that we have measured the same effective temperature
$T_{eff}$ perturbing different degrees of freedom is remarkable.  It
is known that when different degrees of freedom are coupled they must
reach the same temperature. However to our knowledge this is the first
time that such behavior has been measured in the aging regime of a
finite dimensional system.

It is worth noting that there is strong evidence~\cite{nicodemi} that
the internal degrees of freedom are responsible for a thermodynamic
spin glass transition at a temperature near $T_g$, the structural
glass transition temperature. The constant value of the FDR for the
spin variables indicates that the transition should be typical of
systems with one replica symmetry breaking. If this is confirmed we
have at hand a finite model, with short range interactions, which
presents a discontinuous transition, like that present in mean field
models as the $p$-spin. In fact, at a mean field level, a line of
discontinuous transitions have been found in models closely related to
the FILG~\cite{jef}. More work is needed in order to test this
possibility in the short range model.

\begin{figure}
\centerline{\epsfig{file=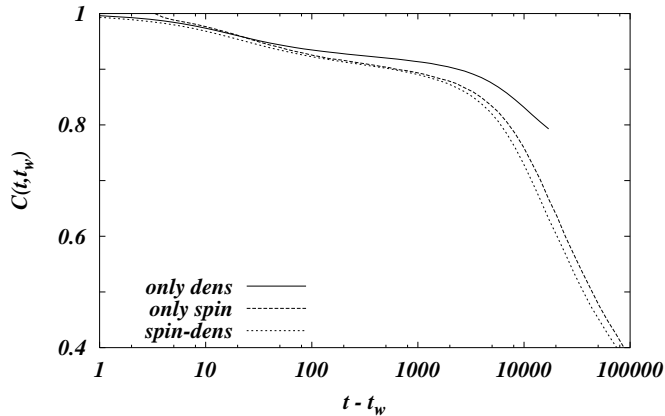,width=\columnwidth,angle=0}}
\caption{Several possible correlations $C_{ns}$, $C_s$ and $C_n$ for
$t_w=10^4$.}
\label{3corr} 
\end{figure}

For the sake of comparison, in Fig.~\ref{3corr} results for all three
correlations studied are shown for $t_w=10^4$. Notice that the
correlation involving both occupation and spin variables is always
smaller than the other two: $C_{ns} < min(C_n,C_s)$. For small times,
$C_{ns}$ is dominated by the particles dynamics, while, for large
times, by the internal degrees of freedom dynamics.

\section{Conclusions}
\label{section.conclusions}

We have described the main characteristics of the out of equilibrium
dynamics of the three dimensional Frustrated Ising Lattice Gas. It
shows a very complex and interesting phenomenology similar to that
observed in real glass formers. The glass transition seems to have a
purely dynamical origin as no evidence of singular behavior was found
neither in the linear nor in nonlinear compressibilities.  Aging
dynamics is present in both density and internal degrees of freedom
and the decay of autocorrelations is similar to that observed in
molecular dynamics simulations of Lennard-Jones glasses. Cage effects
are evident both in density relaxation and on diffusion properties
where a clear characterization of different regimes is possible. From
slow compression experiments different densities are reached depending
on the compression rate and hysteresis is observed from cycling
experiments. After a sudden compression the density slowly approaches
a limiting value. Aging experiments at fixed densities slightly above
and below this limiting value show clear qualitative differences
showing that this density may separate two different dynamical
regimes.

The overall scenario is very reminiscent of mean field scenarios for
the glass transition, except for the presence of compression rate
dependence which is absent in mean field. Interestingly, the form of
the fluctuation-dissipation theorem violations is exactly that
predicted in mean field approximations, i.e.\ a constant
fluctuation-dissipation ratio . This means that this model presents
two very separated time scales.  The effective temperature in the out
of equilibrium scale is nearly temperature independent in the glassy
phase. This raises the question whether activated processes play or
not an important role in its dynamics.  The same constant effective
temperature is observed associated with the internal degrees of
freedom. At mean field level, a constant effective temperature has
been obtained in systems characterized by one step replica symmetry
braking (1RSB). Thermodynamically, a phase transition with a divergent
susceptibility associated with internal degrees of freedom is
observed.  This may be the first realization of a glassy phase
characterized by 1RSB in a finite dimensional model. The presence of a
static transition associated with density variables at a low
``Kauzmann temperature'' $T_K < T_g$ remains an open problem in this
model. It would be interesting to study in more detail how the
restoration of FDT relations is achieved as $T_g$ is approached from
below in order to test the strength of the time scales
separation. Also activated processes may show up for bath temperatures
larger than those explored in these simulations as long as frustration
begins to be lifted by thermal activation.

F. R.-T. thanks S. Franz for useful discussions on the subject.  This
work was partly supported by the Brazilian agency CNPq. JJA
acknowledges the Abdus Salam International Centre of Theoretical
Physics for support during his stay, where part of this work was done.

\end{multicols}

\begin{thebibliography}{30}
\bibitem{note1} In the presence of a field the perturbation is
actually the {\it lack} of magnetization, which in fact grows with
time toward the ``right'' value.
\end{thebibliography}
\end{document}